\documentclass[structabstract]{aa}  
\usepackage{graphicx}
\usepackage{txfonts}
\usepackage{natbib}
\usepackage{amssymb}
\begin{document}
   \title{An occultation event in the nucleus of the planetary nebula M 2-29
          }

   \subtitle{}

   \author{M. Hajduk\inst{1} \and A. A. Zijlstra\inst{2}
           \and K. Gesicki\inst{1}
	  }


   \institute{Centrum Astronomii UMK, 
              ul.Gagarina 11, 
	      PL-87-100 Torun, 
	      Poland
              \\
              \email{Marcin.Hajduk@astri.uni.torun.pl,
	      Krzysztof.Gesicki@astri.uni.torun.pl
	      }
              \and
	      University of Manchester,
              School of Physics \&\ Astronomy,
              Oxford Street,
              Manchester M13 9PL, UK \\
              \email{a.zijlstra@manchester.ac.uk}
             }


  \abstract
   {}
   {Eclipses and occultations of post-AGB stars provide a powerful method
   of exploring the near-stellar environment, including close companions
   and  circumstellar debris disks. Only six eclipsing systems and one
   dust-occultation system are currently known. New cases are important
   for our understanding of binary evolution during the AGB mass-loss
   phase.}
   {We study the post-AGB central star of the (bipolar) Galactic bulge
   planetary nebula M\,2-29. We have obtained additional HST imaging and
   SAAO spectroscopy of the object.}
   {The star showed a pronounced, long-lasting occultation with
   subsequent recovery. The event lasted almost 3 years, with a
   secondary minimum 9 years later. The photometric behavior of M\,2-29
   resembles the dust-occultation events seen in NGC\,2346, and is
   modeled as an occultation by a circumbinary disk, where the binary
   period is 18 yr. Modulation during the decline shows evidence of
   another companion with a period of 23 days.}
    {M\,2-29 is the first eclipsing disk system among post-AGB stars.
    Close binaries with periods of around 1 month, as found in M\,2-29,
    have been proposed to supply the energy needed to create the tori of
    bipolar planetary nebulae.}


   \maketitle
%

\section{Introduction}

The role of binary stars in the formation and evolution of planetary
nebulae (PNe) is a controversial one \citep{Zijlstra2007}. In the
traditional view, the ejection of the nebula occurs through a
radiation-driven superwind, at the tip of the asymptotic giant branch
(AGB); however, the outflow may instead be driven via an interacting
binary system, providing a source of angular momentum to the outflow
\citep[e.g.][]{Livio1988}. Even if not directly interacting, binary
companions can trap some previously ejected material into disks, as
commonly seen around cool (A-F)post-AGB stars \citep{vanWinckel2003} with
orbital periods of 100--1000 days.

Information on binary systems and disks within PNe is limited.  Only
about 20 compact binary systems are known, and all but one have periods
in the range 0.1--5 day \citep{marco2008}. Six eclipsing systems are
currently known \citep{Miszalski08, marco2008}, and they provide valuable
but rare information on close companions. Their periods are between 0.17
and 4.9 days.

A few PNe show evidence of trapped circumstellar material. The VLT has
detected a 10\,AU disk around the central star of the Ant nebula
\citep{Chesneau2007}. Infrequent obscurations by orbiting dust clouds
provide evidence of a similar structure around the central star of
NGC\,2346 (V651 Mon) \citep{mendez}. NGC\,2346 is a known binary system.
With a period of 16 days it is so far the only known wider system, with a
period some 5 times longer than the next longest. It approaches the
period range seen for the A--F post-AGB stars with disks.

\citet{Peretto2007} suggest that NGC\,6302 has an undetected binary
companion with a period of about 1 month. Such a binary could have
supplied the energy necessary for its expanding, high-mass torus. If this
is correct, the implication is that PNe with dense tori have binaries
with similar periods and possibly central remnant disks. 

\begin{figure}
   \centering
   \includegraphics[width=88mm]{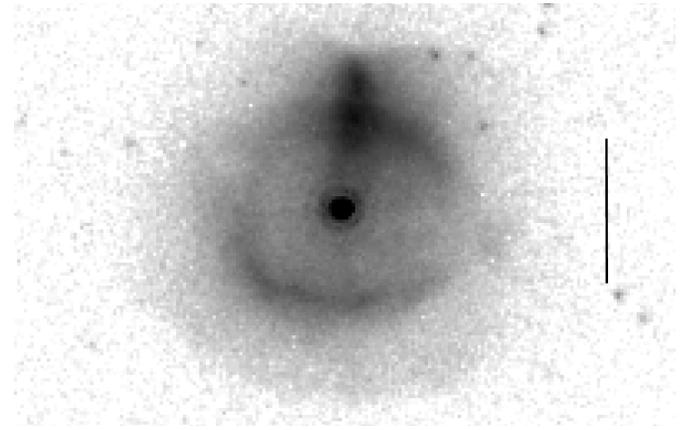}
      \caption{H$ \alpha$ HST image of M\,2-29 obtained in 2003. A
      logarithmic intensity scale is applied. The nebula is complex,
      including a ring, a halo, a jet-like structure, and a central star.
      North is on top, east to the left. The length of the bar is 2
      arcsec.
              }
    \label{Halpha_image}
\end{figure}

The planetary nebula M\,2-29 (PN\,G\,004.0$-$03.0) is a 4-arcsec object,
previously considered as a halo PN \citep{pena}, but the revised
abundance analysis of \citet{torres} suggests it to be a member of the
bulge. Old HST images are described by \citet{torres}. New HST Wide
Field Planetary Camera 2 images in [O\,{\sc iii}] 5007 {\AA} and $\rm H
\alpha$ lines and in continuum were obtained on May 20, 2003. The
H$\alpha$ image (Fig. \ref{Halpha_image}) confirms the features described
by \citet{torres} but at much better sensitivity. (The jet-like feature
will be discussed elsewhere: we attribute it to interaction with the
interstellar medium  \citep{Wareing}.) The 2MASS NIR colors of $J-H =
0.41$ and $H-K = 1.17$ differ from those of ionized gas, and are
indicative of a hot star plus a 1000\,K black body \citep{Whitelock}.

In this paper we present data on a new occultation event in a central
star of a planetary nebula, lasting several years.

\section{Observations and data reduction}

\begin{figure}
   \centering
   \includegraphics[width=88mm]{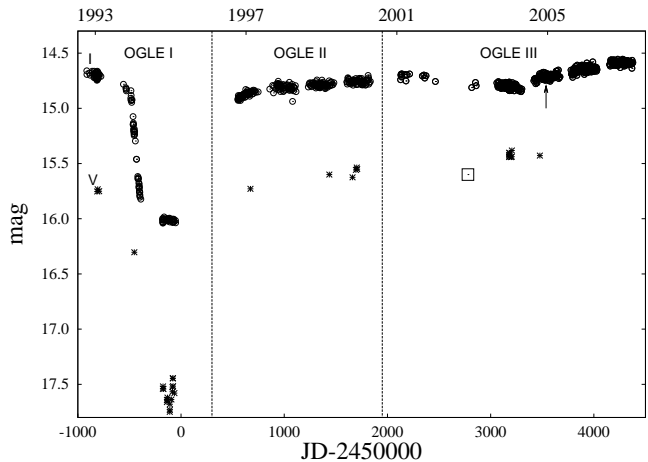}
      \caption{$I$ and $V$ lightcurve of M\,2-29 from OGLE. The rectangle
      gives the stellar magnitude obtained from the HST 547 nm image. The
      arrow marks the SAAO spectroscopic observation.
              }
         \label{Ogle}
\end{figure}

M\,2-29 has been observed by the OGLE (Optical Gravitational Lensing
Experiment) survey since 1992.  OGLE observed the star regularly in the
$I$ band and occasionally in the $V$ band. The central star of M\,2-29
is found to exhibit unusual light variations. \cite{zebrun} lists it as
an R\,CrB-type star candidate. The lightcurve of the object presented in
Fig.\ref{Ogle} shows an eclipse-like event, lasting several years (many
times longer than the obscuration events in RCrB stars). Out of minimum,
the lightcurve shows other slow changes.

The OGLE data covers three phases of the survey: OGLE-I (1993-1995),
OGLE-II (1997-2000) and OGLE-III (2001 onward), over a total period of 14
years, with short periods of avoidance each year. OGLE-I operated at the
1-m Swope telescope at the Las Campanas Observatory. After a one-year
break in 1996, the project moved to the dedicated 1.3-m Warsaw Telescope.
OGLE-II was performed using a single SITe CCD camera, while the third
phase of the project used eight binned CCD chips.

Only the OGLE epoch-II photometry is calibrated in absolute units. An
average systematic shift of $-0.04$ and $+0.26$\,mag at $I$ band (with an
RMS of 0.03) was applied for OGLE-I and OGLE-III, respectively, based on
comparing the magnitudes of 8 field stars. The $I$-band pipeline
photometry uses the DIA (Difference Image Analysis) method, where a
reference image is subtracted before the photometry is carried out. The
$V$-band pipeline photometry is done differently, using DAOPHOT
PSF-fitting. To obtain correct stellar magnitudes, we measured the $V$
band magnitudes again with the DIA method, subtracting the reference
image from other images.

\begin{figure}
   \centering
   \includegraphics[width=88mm]{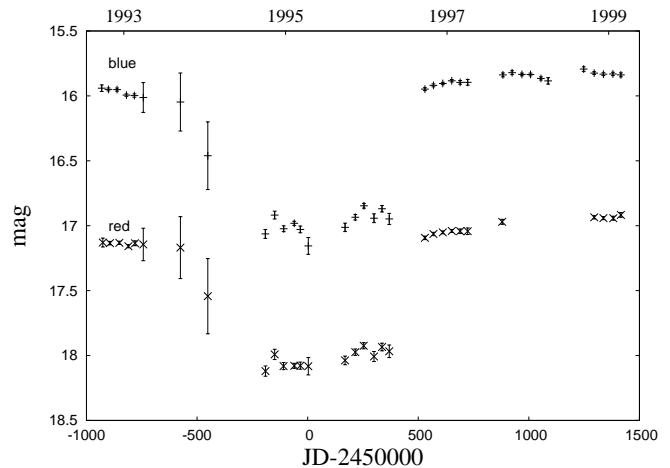}
      \caption{Blue and red lightcurve of M\,2-29 from MACHO, averaged
      over 40-day bins.
              }
         \label{Macho}
   \end{figure}
  
\begin{figure}
   \centering
   \includegraphics[width=88mm]{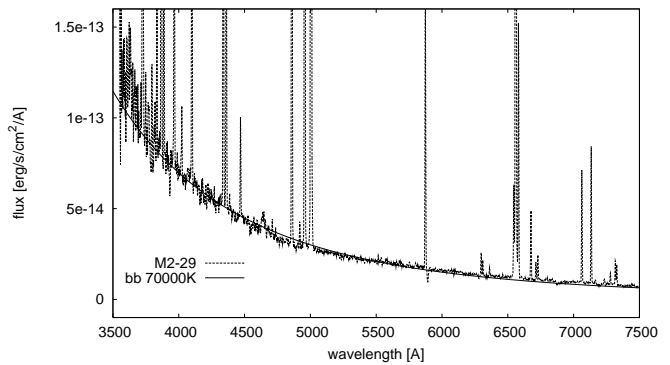}
      \caption{Dereddened SAAO spectrum of M\,2-29 along with the
      black-body 70\,kK model.
              }
         \label{Spectrum}
\end{figure}
  
The OGLE $V$ and $I$ filters are close to the Landolt standards
\citep{udalski}.  The $I$ bandpass avoids strong nebular emission lines
and is dominated by the stellar emission. This is confirmed by the
stellar appearance of M\,2-29 in the $I$ images. The $V$-image is
affected by the nebular [O\,{\sc iii}] 5007 \AA\ line, located at the
blue edge of the filter. The $V$ band OGLE images indeed show a barely
resolved elongated structure superposed on a stellar PSF in the center of
the nebula. The sensitivity of the system at 5007 \AA\ varies between the
three epochs, which causes systematic offsets in the determination of the
stellar brightness. The measured stellar magnitude is very sensitive to
the nebular contribution, especially during  minimum when the bulk of the
measured $V$ band flux comes from the nebular emission. To remove the
nebular contribution and obtain absolute photometry, we smoothed the HST
[O\,{\sc iii}] 5007 \AA\ image down to the resolution of the reference
image and subtracted it from the reference frame. A scaling factor was
chosen such that, after subtraction, a pure stellar PSF was left.

The contribution of the emission line to the OGLE V-band photometry was
checked with the help of a 1.9-m SAAO telescope low-dispersion optical
spectrum obtained on June 16, 2005. The low-resolution (5\,\AA) spectrum
covers the wavelength range 3500--7500 \AA.

MACHO data is available for M\,2-29 for 7 seasons, starting from 1993
(Fig. \ref{Macho}). The blue MACHO filter runs from $\sim$4500 to 6300
\AA, including the [O\,{\sc iii}] 4959/5007 \AA\ line. The red filter
($\sim$6300--7600 \AA) includes $\rm H \alpha$ \citep{alcock}. The MACHO
photometry is not calibrated to absolute units. To improve the
S/N, the data points were averaged for each 40 days.

The SAAO spectrum (Fig. \ref{Spectrum}), similar to the spectra
published by \cite{pena}, shows a near-featureless stellar continuum. 
The He\,{\sc ii} 4541\AA\ absorption line \citep[reported by][]{pena2}
is confirmed in our spectrum.  We dereddened the spectrum of M\,2-29
using a standard extinction law, with $\rm c_{\beta}$ of 1.18,
corresponding to $E(B-V)=0.63$, and $A_V = 1.9$. There is no indication
for a binary companion in the spectrum. Our photo-ionization model of
the nebula indicates a stellar temperature $T_{\rm eff}\approx 70\,$kK,
$\log (L/L_\odot) = 3.4$ and an ionized mass $M_i \sim 0.48\,M_\odot$,
for a distance of 8\,kpc.  The radius of the central star is about $\rm
R_{\star} = 2 \times 10^{10} cm$.

\section{Discussion}

\subsection{The lightcurve}

\begin{figure}
   \centering
   \includegraphics[width=88mm]{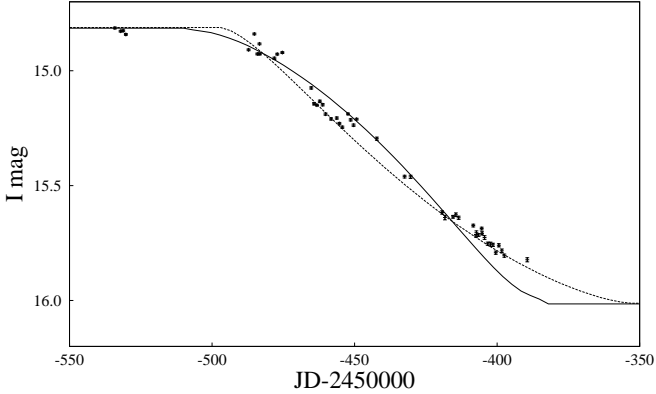}
   \caption{$I$ light curve of M\,2-29 from OGLE during the decline. 
   Two models are shown for an eclipse by the object by a sharp-edged
   screen. Dashed line: a total eclipse of the star + constant source of
   the flux present (or an eclipse by a semi-transparent cloud). Solid 
   line: partial eclipse by a rectangular, opaque body.
           }
         \label{Decline}
\end{figure}

Between the beginning of June (JD 2449512) and September 1994 (JD
2449610), for $\sim$100 days, the $I$-band magnitude dropped from $\rm
14.9$ to $\rm 16.0$ mag (Fig. \ref{Ogle}). The drop after JD\,2449512
was approximately linear on the magnitude scale, showing tiny variations
(Fig. \ref{Decline}) superposed on the linear trend. The final part of
the decline was not covered by OGLE. If the drop of the brightness did
continue to be linear, the star would have reached the minimum
approximately at JD 2449620, with the full decline taking about 110
days.  For the next 210 days, it remained at the minimum level of 16.0
mag. The MACHO data indicate that the minimum remained flat even after
the OGLE observations ceased, and the recovery took place somewhere
between JD 2450380 and 2450507 and lasted less than 130 days (Fig.
\ref{Macho}). The whole event took $\approx$ 1000 days.

Out of the minimum, a gradual, linear brightening is seen between 1998
and 2007, by $\Delta I =0.2\,$mag. The continuum was brighter before the
eclipse event, suggesting a reversal of this trend occurred around the
minimum. Superposed on this trend is a secondary minimum, during
2002--2005, with an amplitude of $\Delta I ={\rm 0.15}$. This secondary
minimum is not sampled well regarding its onset. Its duration was
similar to that of the primary minimum, but it lacked the flat minimum.
The two minima are separated by  $\sim 9\,$yr, giving a plausible
orbital period of 18\,yr. Confirmation is needed.

The amplitude of the decline was $\Delta I \sim \rm 1.3$\,mag, and
$\Delta V \sim \rm 1.7 - 1.8$\,mag. MACHO data shows a decline of $\sim
\rm 1.0$\,mag in MACHO blue and $\sim \rm 0.9$\,mag in MACHO red. The
star may have shown somewhat redder colors during minimum than during
maximum.

\subsection{Disk occultation}

The notable aspects of the eclipse are the long time scale, the slow
ingress and egress, and the relatively shallow depth. 

An eclipse by a companion star can be ruled out. The decline took about
100 days, and the whole minimum lasted about 1000 days. This implied a
size of the occulting body of $\sim 10\times$ larger than the central 
star, or $\rm 2 \times 10^{11} cm$ ($\sim 3 \, R_{\odot}$), comparable 
to the size of a main-sequence star. However, the relative tangential
velocities would be close to $\rm 2 \times R_\star / 100 d = 25
m/s$, corresponding to an orbit of $\sim 10^6\,$AU.  The eclipsing body
could be on a very eccentric orbit, or not bounded gravitationally with
the central star; however, this does not change the extreme unlikelihood
of such an eclipse event. 

The event is fitted better with an occultation by a large screen. The
central star of a bulge PN is expected to have a mass $M_{\rm f} \approx
0.6\, M_\odot$. Assuming a secondary component with similar mass, the
orbital separation for an 18-yr period is $\sim 7\,\rm AU$. The eclipse
lasted for about 15\%\ of the adopted orbital period: this yields a
radius of the occulting body of $R \sim 1 \hbox{--}5\,\rm AU$, depending
on mass ratio and where the occultation occurred on the orbit.

We propose a model where the occultation is caused by a circumbinary
(dust) disk. Such disks are now known to be common around post-AGB
stars, and especially so around RV Tau stars \citep{vanWinckel2003,
Deroo2007, deRuyter2006}. Their absence around central stars of PNe has
become puzzling. In recent years, the first evidence of such disks
have begun to appear, e.g. the discovery of an unexpected compact disk
in the Ant nebula \citep{Chesneau2007}, a disk in the helix nebula
\citep{Su2007}, and the possibly related structure in CK Vul
\citep{Hajduk2007}. The NIR colors indicate the presence of 1000\,K
dust in M\,2-29, as mentioned above, i.e. near the sublimation
temperature. The dust sublimation radius is around 10\,AU, and so the
dust is located outside of the orbit of the 18-yr binary.

Among objects with similar events, long-lasting minima are observed in
wide binary systems, like symbiotic stars; they are ascribed to dusty
clouds. In V\,Hya such a minimum is observed repeatedly with over a 6000
day period \citep{Knapp1999}. NGC\,2346 has shown two long-lasting
eclipses by clouds \citep{Kato2001}. Either $\epsilon$ Aur or EE\,Cep are
examples of eclipses by a disk placed nearly edge-on to the observer,
around an unseen companion \citep{carroll, mikolajewski}. Such eclipses
appear to be flat and almost grey.

Many of the obscuration characteristics of M\,2-29 are replicated in the
`winking' star KH 15D, an eccentric pre-main-sequence binary that is
gradually being occulted by an opaque screen \citep{Winn, Herbst2008}. 
The binary orbit is inclined with respect to the circumbinary disk, and
in the current configuration one of the stars experiences a `sunset' and
`sunrise' every orbit behind the disk. The lightcurve of KH 15D is very
similar to that of M\,2-29, although the time scales are very different
as KH 15D presents a much more compact system.

The nature of the secondary star in M\,2-29 is uncertain. We find no
evidence that is has been observed. The depth of the secondary minimum
corresponds to a flux only a few times less at $I$ than the PN central
star, which is much brighter than any plausible main-sequence star. Such
a luminous (red) star would have shown up in the spectra. It is possible
that the secondary minimum, and in fact the slow brightening prior, are
caused by reflected light from the dusty disk. This would explain the
lack of spectral features, the equal length of the secondary minimum
(implying a similar orbital radius), and the partial nature of the
secondary eclipse, which shows that the secondary emission region is
larger than the primary star.

\subsection{Ionized gas}

In the HST images, the star appears much brighter in F656N ($\rm H
\alpha$) than expected from the F547M continuum image: $V=15.6 \pm 0.1$
versus $m({\rm 656N}) \approx 12.9$, corresponding to an H$\alpha$ EW of
$\approx$ 250 \AA. The stellar magnitude was measured by conversion of
the data count from the HST image into the flux and subsequently turned
into Johnson magnitudes using Bessel zero points. \cite{torres} also
finds that the star is blended with an unresolved, inner emission-line
nebula. There may also be an excess in F502N ([O\,{\sc iii}]), where we
estimate $m \sim 14.3$, which after correction for extinction of
$E(B-V)=0.63$ leaves about 1\,mag excess. The [\,O\,{\sc iii}] 5007
\AA\ excess can only arise from low-density gas and argues against a
stellar wind as the origin of the excess. Instead a circumstellar
emission region is indicated.

These data suggest that the light source may be an ionized, inner
component to the disk. Such a disk can cause the stellar excess [O\,{\sc
iii}] and H$\alpha$ emission \citep{Soker2006}. We also tentatively
assign the He\,{\sc ii} absorption seen by \citet{pena2} to this
disk/component. As the ionized region would be much larger than the
stellar radius, this might also help explain the long duration before
reaching the minimum.

\subsection{A close binary}

\begin{figure}
   \centering
   \includegraphics[width=88mm]{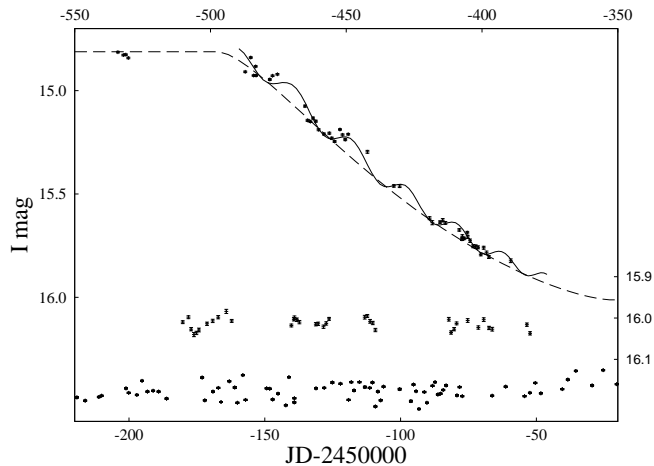}
      \caption{$I$ light curve of M\,2-29 from OGLE during the decline
      and the minimum. The timescale for points from minimum is shifted
      to position the data below the decline phase. A sample of 
      variability of the central star out of the minimum is shown at the
      bottom. The dashed line shows a total eclipse of the central star 
      by a semi-transparent, sharp-edged body with the sharp edge. The 
      solid line includes the $\sim$ 23-day and 18-yr orbital motions of 
      the central star.
              }
         \label{zacmienie}
\end{figure}

We find evidence of semi-regular photometric variations on a timescale of
$\sim$23 days, which appeared during the decline and the minimum and were
present at a level of $\Delta {\rm m}_I \approx 0.05-0.1$\,mag. The
amplitude of these variations is a few times greater than the  accuracy
of the individual data points, which spans $0.004-0.01$ mag. This is
shown in Fig. \ref{zacmienie}: the initially regular variations lost
phase stability before the star was fully hidden. A similar modulation
during the decline was seen for NGC\,2346 \citep{mendez} where the 16-day
photometric variations ceased immediately after the minimum. They were
attributed to the geometric motion of the primary star due to an
undetected, faint companion. As the star moves around the barycenter, the
degree of obscuration varies. This gives a photometric modulation on the
orbital time scale, where the phase depends on the structure of the
occulting material.

Following the case of NGC\,2346, we interpret the photometric variability
as evidence of a companion to the central star with an orbital period of
23 days. The eclipse may be considered as the result of the orbital
motion of the central star behind the obscuring screen. The 18-yr orbital
motion is responsible for the overall shape of the decline, which lasts
for over 2\% of the orbital period. The amplitude and period of the
photometric modulations of the eclipse can be modeled by an orbital
motion of the central star on a $\sim$23-day orbit (Fig. 
\ref{zacmienie}). During each 23-day cycle, there is a phase of nearly
constant luminosity, when the velocity vectors cancel each other and the
star remains stationary with respect to the obscuring body. Assuming that
the outer companion has a mass of $0.5-1 M_\odot$ (as expected for bulge
star) and that the screen is stationary with respect of the barycenter,
this indicates a mass of the close companion of $0.1-0.2 M_\odot$.

In M\,2-29, the photometric variability re-appeared during the minimum.
The phase of the photometric variations is conserved only for 2-3
consecutive periods. This indicates that the optical depth through the
eclipsing body is not uniform. Some of the received light may be
scattered.

The detection of such a companion to the primary central star is 
significant. As shown by \citet{Peretto2007}, the infall of a distant
companion to such a period can provide the energy needed to create an
expanding torus. The morphology of M\,2-29 clearly shows the existence
of such a torus. In contrast, the much shorter-period binaries that
have come through a common-envelope evolution show rings, jets, and
irregular structures, but no tori or butterfly morphologies
\citep{Zijlstra2007}. This would therefore greatly strengthen the case
that close binaries have a strong effect on the shaping of planetary
nebulae.

\begin{acknowledgements}
We acknowledge Prof. Andrzej Udalski and Prof. Michal Szymanski for 
providing the OGLE data. MH thanks S{\l}awek G\'orny for travel to SAAO 
and help in retrieving the spectrum. This work was financially supported 
by MNiSW of Poland through grant No. N\,203\,024\,31/3879.
\end{acknowledgements}

\end{document}